\documentclass[sigconf]{acmart}
\setcopyright{none}
\renewcommand\footnotetextcopyrightpermission[1]{}
\usepackage{footnote}
\usepackage{xcolor}
\usepackage{tikz}
\newfloat{lstfloat}{htbp}{lop}
\floatname{lstfloat}{Listing}

\AtBeginDocument{%
  }

\setcopyright{acmcopyright}
\copyrightyear{2023}
\acmYear{2023}
\setcopyright{acmlicensed}\acmConference[SC-W 2023]{Workshops of The International Conference on High Performance Computing, Network, Storage, and Analysis}{November 12--17, 2023}{Denver, CO, USA}
\acmBooktitle{Workshops of The International Conference on High Performance Computing, Network, Storage, and Analysis (SC-W 2023), November 12--17, 2023, Denver, CO, USA}
\acmPrice{15.00}
\acmDOI{10.1145/3624062.3624543}
\acmISBN{979-8-4007-0785-8/23/11}

\usepackage{booktabs,siunitx}
\usepackage{subcaption}
\usepackage{graphicx}
\usepackage{dirtree}

\usepackage{listings}
\newfloat{lstfloat}{htbp}{lop}
\floatname{lstfloat}{Listing}

\lstdefinelanguage{mlir}{
    alsodigit = {.},
    keywords = {stencil.apply, stencil.access, arith.constant, arith.addf, arith.mulf, stencil.return, f32, f64, i32, i1, stencil.index,stencil.temp,stencil.store,stencil.load,hls.axi_protocol, hls.streamtype,hls.interface,hls.pipeline,hls.unroll,hls.array_partition,hls.dataflow,hls.create_stream,hls.read,hls.write,hls.empty,hls.full,scf.for,func.call}
}

\newcommand\copyrighttext{%
  \footnotesize \textcopyright Gabriel Rodriguez-Canal 2023. This is the author's version of the work. It is posted here for your personal use. Not for redistribution. The definitive version was published in ACM Workshops of The International Conference on High Performance Computing, Network, Storage, and Analysis (SC-W 2023), https://doi.org/10.1145/3624062.3624543"}
\newcommand\copyrightnotice{%
\begin{tikzpicture}[remember picture,overlay]
\node[anchor=south,yshift=10pt] at (current page.south) {\fbox{\parbox{\dimexpr\textwidth-\fboxsep-\fboxrule\relax}{\copyrighttext}}};
\end{tikzpicture}%
}

\begin{document}

\title{Stencil-HMLS: A multi-layered approach to the automatic optimisation of stencil codes on FPGA}

\author{Gabriel Rodriguez-Canal}
\email{gabriel.rodcanal@ed.ac.uk}
\orcid{1234-5678-9012}
\affiliation{%
  \institution{EPCC, \\The University of Edinburgh}
  \streetaddress{47 Potterrow}
  \city{Edinburgh}
  \country{United Kigndom}
  \postcode{EH8 9BT}
}

\author{Nick Brown}
\email{n.brown@epcc.ed.ac.uk}
\affiliation{%
  \institution{EPCC, \\The University of Edinburgh}
  \streetaddress{47 Potterrow}
  \city{Edinburgh}
  \country{United Kingdom}
  \postcode{EH8 9BT}
}

\author{Maurice Jamieson}
\email{maurice.jamieson@epcc.ed.ac.uk}
\affiliation{%
  \institution{EPCC, \\The University of Edinburgh}
  \streetaddress{47 Potterrow}
  \city{Edinburgh}
  \country{United Kingdom}
  \postcode{EH8 9BT}
}

\author{Emilien Bauer}
\email{emilien.bauer@ed.ac.uk}
\affiliation{%
 \institution{School of Informatics, \\The University of Edinburgh}
 \streetaddress{Informatics Forum, 10 Crichton St, Newington}
 \city{Edinburgh}
 \country{United Kingdom}}

\author{Anton Lydike}
\email{anton.lydike@ed.ac.uk}
\affiliation{%
  \institution{School of Informatics, \\The University of Edinburgh}
 \streetaddress{Informatics Forum, 10 Crichton St, Newington}
  \city{Edinburgh}
  \country{United Kingdom}}

\author{Tobias Grosser}
\email{tobias.grosser@ed.ac.uk}
\affiliation{%
  \institution{School of Informatics, \\The University of Edinburgh}
 \streetaddress{Informatics Forum, 10 Crichton St, Newington}
  \city{Edinburgh}
  \country{United Kingdom}
  \postcode{78229}}

\renewcommand{\shortauthors}{Rodriguez-Canal et al.}

\begin{abstract}
The challenges associated with effectively programming FPGAs have been a major blocker in popularising reconfigurable architectures for HPC workloads. However new compiler technologies, such as MLIR, are providing new capabilities which potentially deliver the ability to extract domain specific information and drive automatic structuring of codes for FPGAs.

In this paper we explore domain specific optimisations for stencils, a fundamental access pattern in scientific computing, to obtain high performance on FPGAs via automated code structuring. We propose Stencil-HMLS, a multi-layered approach to automatic optimisation of stencil codes and introduce the \textit{HLS} dialect, which brings FPGA programming into the MLIR ecosystem. Using the PSyclone Fortran DSL, we demonstrate an improvement of 14-100$\times$ with respect to the next best performant state-of-the-art tool. Furthermore, our approach is 14-92$\times$ more energy efficient than the next most energy efficient approach.
\end{abstract}



\maketitle
\pagestyle{plain}
\copyrightnotice

\section{Introduction}

Programming FPGAs is hard, and it is even more difficult to program these effectively to exploit the full potential of the hardware. Arguably this is one the main reasons that the technology has not become more commonplace in our supercomputers, As the supercomputing community continues to progress into the exascale era, and our HPC machines become more heterogeneous, it is not realistic to expect scientific programmers to be able to master all these complex, architecture specific, performance optimisation techniques. One solution is that of Domain Specific Languages (DSLs), which are high-level abstractions that provide a richness of expressivity which enables the scientific end users to express their problem in a manner that is far closer to their domain of expertise. The term \emph{language} is somewhat of a misnomer as instead it has become popular for DSLs to provide abstractions within existing programming languages that provide an interface for expressing a rich amount of information about the problem that can be passed down the compiler stack for the tooling to make informed, automatic, decisions around optimisation. 

Being able to effectively exploit FPGAs for high performance workloads is a prime example of where a DSL is potentially helpful. Whilst there are numerous studies of HPC applications on FPGAs, these involve significant manual optimisation of the code \cite{klaisoongnoen2022fast} \cite{karp2021high} \cite{brown2021porting}. Not only is this a time consuming process that requires significant FPGA expertise from the programmer, but furthermore the resulting code is often very different from its CPU or GPU counterpart. An example of this type of problem is with stencil-based kernels, which underlie many scientific codes. Stencils are ubiquitous in HPC applications, and whilst there have been numerous studies exploring the techniques required in gaining optimal performance on FPGAs \cite{de2018designing}, it is not realistic to expect scientific programmers to be able to master these.

In this paper we use PSyclone, which is a Fortran based DSL for weather and climate simulations, as a vehicle to explore our approach in driving automatic optimisations of classes of HPC algorithms via MLIR for FPGAs. Focusing on stencil-based codes due to their ubiquity, we propose Stencil-HMLS (High-Multi Level Synthesis), a framework for automatic optimisation of stencil codes. Hence the contributions of this paper are as follows:
\begin{enumerate}
	\item A new \textit{HLS} MLIR dialect which abstracts dataflow concepts on FPGAs and is able to then be lowered directly to AMD Xilinx's HLS backend.
    \item Transformations from the existing MLIR \textit{stencil} dialect to our HLS dialect, demonstrating that one can leverage the rich amount of domain specific information to apply dataflow specific optimisations.	
	\item Demonstration of how MLIR can be integrated with DSLs such as PSyclone to target FPGAs in a manner that requires little effort on behalf of the DSL developer. 
\end{enumerate}

The rest of this paper is structured as follows: Section \ref{sect:related_work} provides an outline of the related work and establishes comparison points with our work that we will pick on on the evaluation, Section \ref{sect:implementation} describes the implementation of the HLS dialect and the transformations that enable the automatic optimisation of stencils on FPGAs, Section \ref{sect:evaluation} presents the results that assess the performance of Stencil-HMLS and compares it against the state-of-the-art. Finally, Section \ref{sect:conclusions} showcases the conclusions and pointers for further work.

\section{Background and related work} \label{sect:related_work}

\subsection{Auto-optimisation of codes for FPGAs}
There have been several efforts to synthesise Von-Neumann based codes which were not designed for dataflow architectures into a form that will deliver good performance on FPGAs. Some approaches, such as DaCe \cite{ben2019stateful}, advocate performance portability, where the framework compiles codes into a dataflow IR, that they call Stateful Dataflow MultiGraph (SDFG), and this is then lowered to CPU, GPU or FPGA targets. DaCe is interesting because it  seeks to decouple the responsibilities of the domain scientist from the performance specialist, where the first can focus on their domain of mathematics using Python that is then lowered into an SDFG representation which can be optimised either manually by the performance engineer or automatically by the tooling. In order to achieve good results, however, this approach requires either manual construction of the SDFG or significant optimisation by the performance expert of the generated SDFG. 

The StencilFlow \cite{de2021stencilflow} tool has been built atop DaCE as it generates SDFGs for stencil based kernels based upon a JSON description. Although it is possible to generate the JSON description automatically, this is highly code specific and there is no general way in which the JSON can be generated without modifying the tool.

Other tools leverage MLIR for the generation of a multi-layered IR that can be optimised on different levels of abstraction and apply optimisations automatically through design space exploration (DSE). This technique relies on heuristics to explore different combinations of parameters of well known techniques, such as the unrolling factor or the size of loop tiling. ScaleHLS \cite{ye2021scalehls} and SODA-opt \cite{agostini2022soda} are both examples of technologies which rely on this principle. 

AMD Xilinx Vitis HLS is a high-level synthesis (HLS) framework that synthesises C/C++ codes into hardware. Driven by code annotations in the form of pragmas and a template library, it follows the principle of minimising the initiation interval (II), i.e. the number of cycles between iterations of a loop, to deliver performance. Vitis HLS provides a complete compilation from the LLVM C++ frontend based on Clang which generates LLVM-IR that is provided to the AMD Xilinx LLVM backend which generates HDL. 

There are other non-vendor specific HLS solutions, such as Bambu HLS \cite{ferrandi2021bambu} which is also able to generate HDL from C/C++, and these follow similar principles to the Vitis HLS frontend. Fortran-HLS \cite{fortran-fpga} couples the LLVM Flang frontend with AMD Xilinx's backend to enable Fortran programming for FPGAs, with the objective of enabling HPC developers to avoid the initial time consuming step of porting their codes into C++ when targeting FPGAs. However, crucially whilst these tools tend to enable correct by construction dataflow codes, the automatic optimisation that they are able to perform is limited and-so significant programmer expertise is required to obtain best performance on the FPGA.




\subsection{MLIR and xDSL}
Multi Level Intermediate Representation (MLIR) is an LLVM sub-project which aims to innovate in the area of Intermediate Representation (IR). For many years LLVM backends, which target specific architectures such as AMD Xilinx FPGAs, CPUs, and GPUs are connected with the language-specific frontends via LLVM-IR. However, LLVM-IR is rather low level and these language frontends, such as Clang, must undertake significant work to transform a programmer's source code into this form. An example is in leveraging FPGAs, where AMD Xilinx have had to modify the Clang front-end to provide support for their pragmas and there is very limited optimisation of the programmer's code for a dataflow architecture.

MLIR looks to redefine the use of IRs by providing a hierarchy of IR dialects and transformations between these. First developed by Google as part of their Tensorflow project, MLIR was released open source in 2019 and became part of LLVM shortly after that. Using MLIR, frontends can generate more suitable, higher level, representations of a user's code before this is progressively lowered through a series of dialects and ultimately LLVM-IR. Crucially, these MLIR dialects and transformations already exist and-so a frontend developer needs to only generate this highest level of MLIR dialect, and can then rely on this being lowered.

MLIR provides a rich set of standard dialects, and using the framework it is possible for new dialects and transformations to be developed. Indeed, AMD Xilinx have developed their own bespoke MLIR dialect for the Versal AIE engines that integrates with their proprietary downstream compiler via lowerings. However, there is no HLS dialect that has been provided so, until this work, there has been no MLIR integration with HLS for programming the PL.

However, a limitation with MLIR is that it is written in complex C++, uses an esoteric description language for definitions of dialects, and is a continually changing target due to the large development community. To this end xDSL has been developed which is a Python based compiler framework. xDSL is fully compatible with MLIR, and IRs can be moved between the two technologies seamlessly. In this paper we develop our dialects and transformations in Python using xDSL as this enables much faster prototyping of our central concepts and ideas, with these then able to be fed back into the main MLIR code-base once they are fully mature.
 
\subsubsection{The MLIR stencil dialect}

The MLIR \emph{stencil} dialect enables the representation of  stencil computations at the IR level. By providing this high level representation of a programmer's mathematical problem, there is then a rich amount of information upon which the compiler has to operate.

\begin{lstfloat}
\begin{lstlisting}[language=mlir, frame=lines, label=lst:stencil_example, numbers=left, caption=Example MLIR for 1-dimensional 3-point stencil calculation]

  %source = stencil.load(%114) : (!field<[0,128]xf64>)
                               -> !temp<?xf64>
  %out = stencil.apply(%arg = %source : !temp<?xf64>)
                    -> !temp<?xf64> {
    %l = stencil.access %arg[-1] : f64    
    %r = stencil.access %arg[1] : f64
    %v = arith.addf %l, %r : f64    
    stencil.return %v : f64
  }
  
  stencil.store %out to %target([1]:[127])
       : !temp<?xf64> to !field<[0,128]xf64>
  
\end{lstlisting}
\end{lstfloat}

A stencil example which sums neighbouring values in 1D is expressed with this dialect in \autoref{lst:stencil_example}. This would be generated by the frontend tooling from the programmer's description, and the \emph{stencil.load} operation at line 1 loads in values into a temporary type that can be operated upon, likewise the \emph{stencil.store} operation at line 11 stores resulting output back into a form that the rest of the IR can leverage. The \emph{stencil.apply} operation, at line 3 defines the stencil computation itself and there are four operations that comprise the calculation. The \emph{stencil.access} operations at lines 5 and 6 load a stencil value with a relative offset to the current index, here the direct neighbours, the \emph{arith.addf} operation at line 7 adds these values together and this is returned for the current grid cell through the \emph{stencil.return} operation. This \emph{stencil.apply} operation defines the calculations that need to be undertaken on a grid cell by grid cell basis and is executed across the entire grid.

Lowerings from the PSyclone and Devito DSLs, as well as from Fortran more widely in Flang, have been developed for the stencil dialect. Furthermore, transformation passes have been developed for lowering the stencil dialect to CPUs, GPUs, and distributed memory parallelism. The programmer using these DSLs, or Flang, is able to therefore target these architectures with no additional support needed by the frontend apart from generating stencil compatible IR. It is our hypothesis that this dialect can also be used to target FPGAs, where examples such as \autoref{lst:stencil_example} contains enough contextual information such that MLIR transformations can then be written which will operate upon it and lower it to a form optimal for the dataflow architecture.


\section{Implementation} \label{sect:implementation}
\begin{figure*}[htb]
	\includegraphics[width=\textwidth]{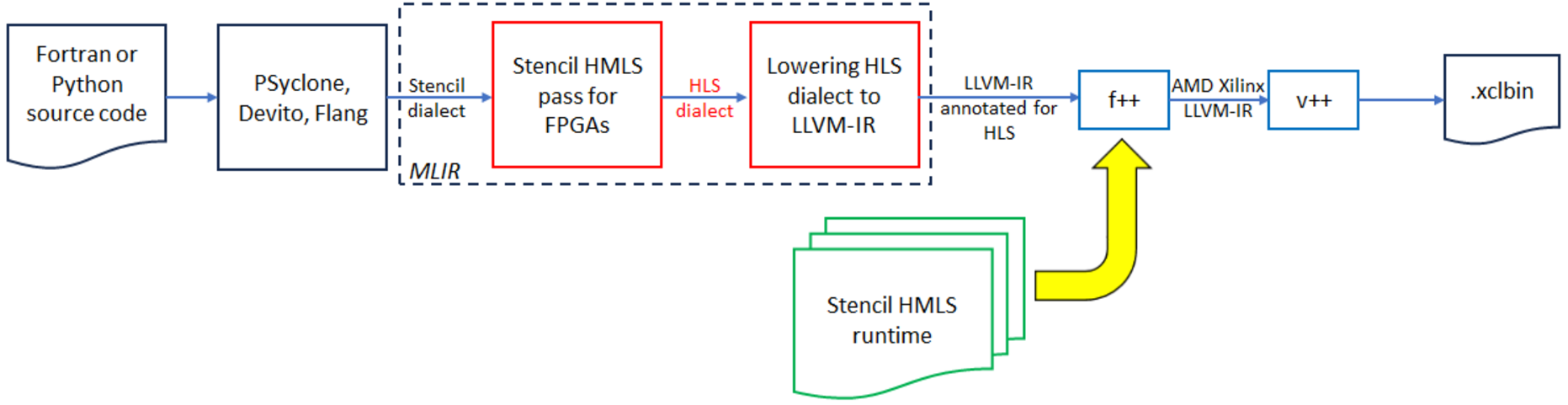}
	\caption{Architectural flow of our approach, where our three main contributions that are discussed in this section are highlighted in red. After source code is converted into the stencil dialect by the DSL or compiler tool, this is then processed by Stencil-HMLS which is manipulated by the \emph{f++} tool which also links against our runtime. The resulting LLVM-IR is provided to the AMD Xilinx HLS backend which generates the HDL.}
	\label{fig:fpp_flow}
\end{figure*}

In this section we explore the steps that we have developed which enable us to automatically lower stencil calculations to optimised dataflow representations on the FPGA. The overall architectural view of our approach is illustrated in Figure \ref{fig:fpp_flow}, where source code is processed by the PSyclone, Devito or Flang tools to generate the stencil dialect. Our transformation then lowers this into our HLS dialect which is lowered into annotated LLVM-IR, and then manipulated by the \emph{f++} tool, which also links against our runtime, and the resulting LLVM-IR is provided into the AMD Xilinx HLS backend. The result is an \emph{.xclbin} file which contains the FPGA configuration. This comprises three main facets, our HLS dialect, the lowering from the stencil dialect to this HLS dialect, and the lowering from the stencil dialect to LLVM-IR which can then be provided to the LLVM backend.

\subsection{HLS dialect}
Our \textit{HLS} dialect replicates the high-level synthesis features provided by the AMD Xilinx Vitis tooling in a vendor agnostic way. Consequently, it is possible to lower the HLS dialect to multiple targets such as LLVM-IR (as we explore here) or CIRCT \cite{eldridge2021mlir}. The HLS dialect currently supports the attributes sketched in Listing \ref{lst:attr} and these describe a specific feature or facet, for instance the \emph{hls.axi\_protocol} is parameterised with the type of AXI protocol which is encoded as an \emph{i32} integer. The attribute \label{attr:attr:streamtype} represents the data type of a stream. 
\subsubsection{Attributes}
\begin{lstfloat}
\begin{lstlisting}[language=mlir, frame=lines, label=lst:attr, numbers=left, caption=Attributes of the HLS dialect]
hls.axi_protocol(%protocol) : (i32) -> hls.axi_protocol 
hls.streamtype(%elem_type) : (f64) -> hls.streamtype 
\end{lstlisting}
\end{lstfloat}

Listing \ref{lst:ops} sketches the operations provided by our HLS dialect, and many of these will be familiar to those who acquainted with Vitis HLS. For instance, the \emph{hls.create\_stream} operation creates an HLS stream of type which is specified as the argument and this is returned as an \emph{hls.streamtype} attribute. The \emph{hls.read}, \emph{hls.write}, \emph{hls.empty}, and \emph{hls.full} operations then operate upon this stream as one would expect who is familiar with Vitis HLS. The \emph{hls.pipeline} operation can be seen, where the desired II is provided as an integer argument, as is \emph{hls.unroll} and \emph{hls.array\_partition} which can be used to optimise for performance.

\subsubsection{Operations}
\begin{lstfloat}
\begin{lstlisting}[language=mlir, frame=lines, label=lst:ops, numbers=left, caption=Operations of the HLS dialect]
hls.interface(%protocol, %bundle) : (hls.axi_protocol, str) -> () 
hls.pipeline(%ii) : (i32) -> ()
hls.unroll(%factor) : (i32) -> ()
hls.array_partition() : () -> ()
hls.dataflow() : () -> ()
hls.create_stream(%elem_type : f64)
hls.read(%stream) : hls.streamtype -> (f64)
hls.write(%stream, %elem) : (hls.streamtype, f64) -> ()
hls.empty(%stream) : (hls.streamtype) -> i1
hls.full(%stream) : (hls.streamtype) -> i1

\end{lstlisting}
\end{lstfloat}

The \emph{hls.dataflow} operation is provided which enables the definition of dataflow regions which are then connected via streams. Whilst this MLIR dialect does not implement the full pragma set of Vitis HLS, it provides enough functionality in these ten operations such that we can leverage them to represent optimised code. 

\subsection{Lowering the HLS dialect to LLVM-IR}
In order to enable the driving of the AMD Xilinx HLS backend we had to develop a transformation which lowered from our HLS dialect to the LLVM-IR. Because existing lowerings exist for the standard MLIR dialects, such as \emph{arith} and \emph{math} to LLVM-IR, then aspects expressed using these, such as mathematical calculations, are already supported by the AMD Xilinx backend. 

Our lowering follows the same approach adopted for connecting Flang with the AMD Xilinx backend \cite{fortran-fpga}, where void functions with no arguments are used to encode HLS directives. The authors of \cite{fortran-fpga} chose to use functions in this manner because they then effectively become annotations in the LLVM-IR and do not alter the structure of the IR. It was found by them to be optimal compared with other approaches which did alter the IR structure and could result in a performance reduction or failure to compile. Once the LLVM-IR is generated, before providing to the AMD Xilinx HLS backend we run the f++ preprocessing tool developed in \cite{fortran-fpga} to identify these corresponding function calls via pattern matching and replace them with the appropriate intrinsics or metadata. In some other cases it is important to take into account the structure of the IR, such as for pipelining or unrolling. These can be applied to any level of nesting in loops and, therefore, f++ makes use of LLVM passes that determine where in the loop tree the call was found to qualify the corresponding body with the necessary metadata.

An added complexity is in the handling of HLS streams, where the AMD Xilinx HLS backend only identifies a legal stream when the following two conditions hold:

\begin{enumerate}
    \item The stream is a pointer to a structure, where the type of the stream will be the type contained within the structure, i.e. the type of the elements read and written from and to the stream will be of this type. For example, the following stream \emph{!llvm.ptr<!llvm.struct<(f64)>>} receives and produces \emph{f64} elements.
    \item There is a call to the intrinsic \emph{@llvm.fpga.set.stream.depth} on the first element of the stream.
\end{enumerate}

To satisfy the first condition we allocate a structure to the type specified in the \emph{hls.create\_stream} operation argument. For the second requirement, we obtain a pointer to the first element of the structure with a \emph{getelementptr} operation using offset \emph{[0,0]}. The first index of this offset specifies the number of top-level structure type elements we are shifting in the pointer, and the second which field we pick in the structure.

\subsection{HLS optimisation passes: stencil dialect to HLS dialect}
\label{sec:stencil-to-hls}
The PSyclone and Devito DSLs both lower into the stencil dialect, and there has been a transformation developed for Flang that will also transform suitable loops into the stencil dialect. These transformations identify appropriate patterns and lower these into the stencil dialect. There is an existing transformation that lowers the stencil dialect into the standard MLIR dialects targeting CPU execution, and initially we added an additional step after this transformation to apply suitable constructs from the HLS dialect so that this could be handled by the AMD Xilinx HLS backend. 

However, this approach was similar to those adopted by the AMD Xilinx Vitis HLS frontend and Bambu HLS, where although the code will execute correctly on the FPGA because it is still structured following the imperative Von Neumann model performance is poor. Instead, because FPGAs benefit from a dataflow approach instead \cite{wang2015hardware} in which they should make progress every cycle, the IR must be organised to suit this architecture.

For stencil computations the most efficient way of organising the dataflow computation is based upon a shift buffer approach \cite{brown2021accelerating} \cite{zhang2007multiwindow}. Using this shift buffer, the stencil operation operates in a sliding window called a shift register instead of the data being directly accessed from external memory on demand, as in the typical CPU implementations. As such, the input field is fetched into the shift buffer until it is full, and then on every cycle required data for the current operation is provided and then a shift operation occurs where each data element is displaced one position to the right. This evicts the oldest element and a new element enters. The shift buffer has found to be an effective optimisation for these types of code, however it adds complexity for the programmer especially when working in three dimensions.

\begin{figure}
	\includegraphics[width=\columnwidth]{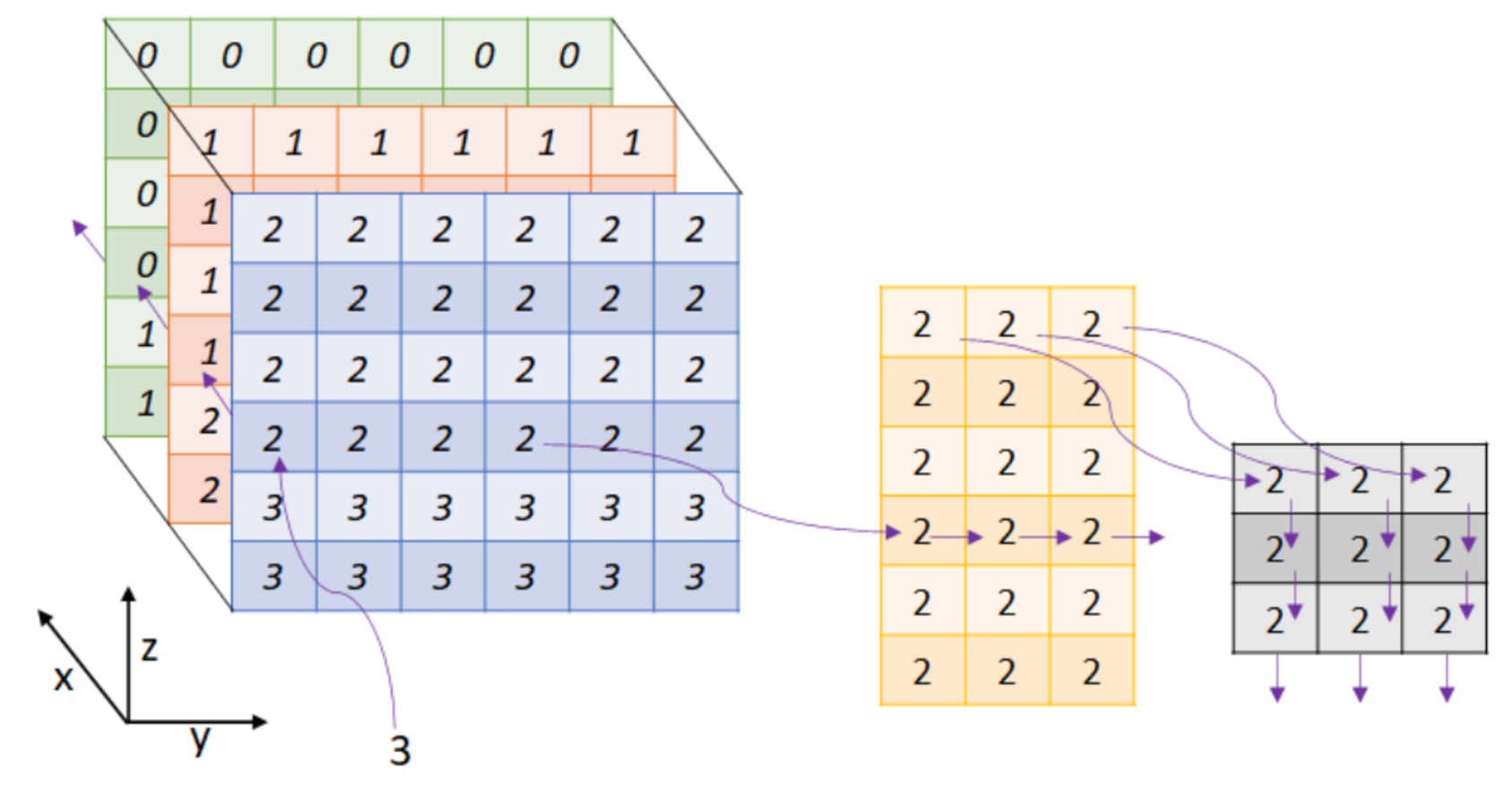}
	\caption{3D shift buffer \cite{brown2021accelerating}.}
	\label{fig:shift_buffer}
\end{figure}


\begin{lstfloat}
\begin{lstlisting}[language=mlir, frame=lines, label=lst:external_load, numbers=left, caption=Lowering of \emph{stencil.external\_load} for the stencil fields.\label{lst:external_load}]
%x_in = hls.create_stream()
hls.dataflow() {
    func.call(%x, %x_in) {"dummy_load_data"}
}
hls.dataflow() {
    func.call(%x_in, %x_shift) {"shift_buffer"}
} 
hls.dataflow() {
// Duplicate streams: 1 copy per field
%x_shift_copy_1 = hls.create_stream
...
%x_shift_copy_ncomp = hls.create_stream
}
hls.dataflow() {
    scf.for(%ncomp) {
        %k = hls.read(%x_shift)
        hls.write(%k, %x_shift_copy_1)
        ...
        hls.write(%k, %x_shift_copy_ncomp)
    }
}

\end{lstlisting}
\end{lstfloat}

\begin{figure}
	\includegraphics[width=\columnwidth]{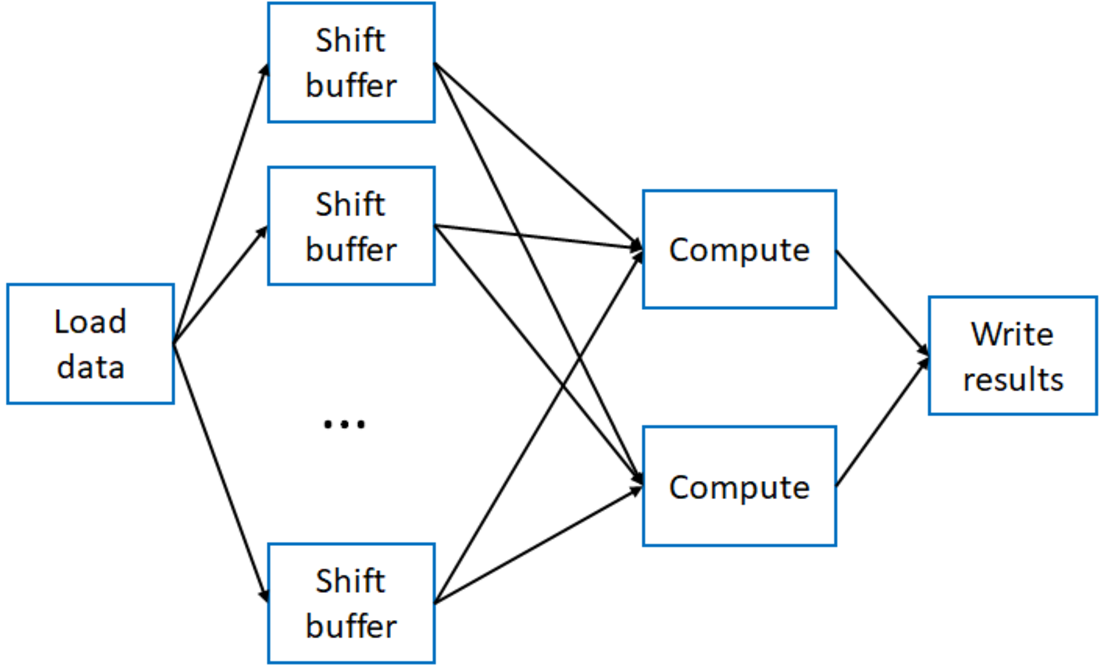}
	\caption{Target dataflow code structure}
	\label{fig:dataflow}
\end{figure}

Our transformation pass that applies automatic optimisation for stencils on FPGAs operates upon the \emph{stencil} dialect IR, like that shown in Listing \ref{lst:stencil_example}, transforming into into a shift buffer pattern using the dataflow approach as illustrated in Figure \ref{fig:dataflow} to ensure that the loading of data, shift buffer(s), computation, and writing of results can operate concurrently for separate elements. Our transformation involves the following steps:
\begin{enumerate}
    \item \textbf{Classification of kernel arguments}: Where the data arguments in a stencil region are classified as either stencil field inputs, stencil field outputs or constants. 

    
    \item \textbf{Replacement of interface type with 512-bit packed version}: To maximise throughput in FPGAs it is important to fetch larger chunks of data, ideally of size 512-bits, each time external memory is accessed \cite{brown2019s}. To that end, the type of the kernel interface's argument pointers is replaced by a 512-packed version of the same type. For example, \emph{f64}, double precision floating point, would be replaced by \emph{!llvm.struct<(!llvm.array<8 x f64>)>}. This new pointer type is then propagated down through the IR to update operations that use it. 

    \item \textbf{Replace direct accesses to external memory by streams}: Moving to a dataflow style of computation is a fundamental step in achieving performance, since it enables us to minimise the II in combination with the shift buffer. The transformation adds a placeholder function, \emph{dummy\_load\_data}, for each stencil array that is being read and generates an HLS stream that will provide elements of data each cycle. It can be seen from Listing \ref{lst:external_load}, that there are three dataflow regions running concurrently, with the first dataflow region calling out dummy function \emph{dummy\_load\_data} and the second region running the shift buffer via calling the function \emph{shift\_buffer}. These operate concurrently and are connected via the \emph{x\_in} stream. The third data flow region in Listing \ref{lst:external_load}, also running concurrently, extracts values from the shift buffer's stream, \emph{x\_shift} and streams these to the compute dataflow stage. As in the illustration of Figure \ref{fig:dataflow}, in the cases where there are multiple compute stages operating upon the same input data then the stream is duplicated in this stage to serve each concurrently. It should be highlighted that our shift buffer does not provide a single value, but instead all the stencil values that could be required. For example, in 1 dimension three values are provided (the current grid cell and two neighbouring cells), in 2 dimensions nine values are provided, and in 3 dimensions 27 values. \label{pass:load_data}, as shown in Figure \ref{fig:shift_buffer}.


    \item \textbf{Separation of stencil fields in the \emph{stencil.apply} operation}: The stencil transformations for the CPU or GPU favour fusing stencils together for fewer, larger stencil regions. However, on the FPGA to obtain optimal throughout it is better for the calculations involved for each stencil field to be split into separate dataflow regions that can run concurrently. Therefore, we identify the result fields through the arguments of the \emph{stencil.return} operation, and for each of the new compute loops we add a \emph{hls.read} operation at the beginning of its body for each of the stencil input fields to read values being streamed from the shift buffer(s). Similarly, an \emph{hls.write} operation is added to the end of the body of each loop to write the result of the resulting field to an output stream. These input and output streams are later connected to the corresponding dataflow stages. 

    \item \textbf{Map \emph{stencil.access} operations to the corresponding stencil value}: The \emph{stencil.access} operation contains an offset of the form \emph{<-1,0,1>} which determines the position of the element in the stencil with respect to the current element in the field. As the shift buffer streams all neighbouring stencil values, this transformation uses that offset to extract the required value and replace the \emph{stencil.access} with the operations required for this.

    \item \textbf{Handle storage of results}: The \emph{stencil.store} operations are replaced by a single call to the \emph{write\_data} dataflow function that writes data to external memory in chunks of 512-bits. 

    \item \textbf{Replacement of placeholder data loading functions} Where the placeholder functions \emph{dummy\_load\_data} from step \autoref{pass:load_data} are replaced by a single call to the \emph{load\_data} function which has been specialised for the number of required input fields. In order to guarantee that data loading occurs before the shift buffer operation, only the first placeholder is replaced and the remaining placeholder function calls are removed. Hence in Figure \ref{fig:dataflow} whilst there are numerous shift buffers, there is only one data loading stage.

    \item \textbf{Copy small data chunks to local FPGA memory}: To minimise accesses to external memory for performance, static data is copied into local FPGA BRAM or URAM if it will fit. After identifying appropriate small data chunks, it is then determined whether this is used in more than one stencil compute loop as if so it must be duplicated to provide the constraint of a single dataflow function accessing a local array. A memory reference from the \emph{memref} dialect is allocated for each of these copies and this is then populated by a loop upon kernel initialisation with static elements from external memory.

    \item \textbf{Assignment of input and output kernel arguments to separate bundles}: To maximise bandwidth between the HLS kernel and external memory the input arguments of the kernel are looped over and the \emph{hls.interface} operation is added into the IR with an AXI4 protocol. Each of these interfaces are assigned to a separate bundle to maximise bandwidth and avoid the bottleneck of a single single physical port which would result in contention as every memory access per cycle would be competing for the same port. The exception to this is for the small data which will be copied to BRAM on kernel initialisation, where this all resides in a single bundle to avoid wasting ports. Each of these ports is connected to a separate bank of HBM.
\end{enumerate}

In the description of the transformation passes described in this section we have made reference to data functions such as \emph{shift\_buffer}, \emph{write\_data} and \emph{load\_data}. These can be found in our runtime which is linked with the generated LLVM-IR at compile time. These functions have been written in C++ and their corresponding LLVM-IR then generated which our tool then links together with the transformed programmer's stencil code.

\section{Evaluation} \label{sect:evaluation}

The assessment of this work has been undertaken using the PSyclone DSL with two real-world 3D stencil kernels. The first is the Piacsek and Williams (PW) advection scheme \cite{piacsek1970conservation}, commonly found in weather simulation codes, such as the Met Office's MONC high-resolution atmospheric model \cite{piacsek1970conservation}. Secondly we evaluate using the tracer advection scheme from the NEMO ocean model which is part of the PSyclone benchmark suite \cite{psyclone_bench}. These two kernels provide different levels of complexity, for instance the PW advection scheme contains three separate stencil computations executing across three fields, whereas the tracer advection kernel comprises 24 stencil computations across six fields. 

We compare our stencil-based approach against three of state-of-the-art HLS tools: DaCe, SODA, AMD Xilinx's Vitis HLS and StencilFlow. We measure performance in million points per second (MPt/s), which is calculated as the size of the problem divided by the execution time of the kernel. The average power draw is measured in watts and calculated based upon the average of the instantaneous power draw of the card over the execution of the kernel. The energy consumed is measured in joules, and calculated as the average power draw times the execution time of the kernel. In gathering the power and energy measurements we followed the method described in \cite{klaisoongnoen2022low}. All results are averaged over 10 runs.

We run our experiments on an AMD Xilinx Alveo U280 FPGA which is being driven by the host containing a Cascade Lake Xeon Platinum 8260M CPU. We aim to maximise the use of the FPGA by replicating the compute units (CU) where possible and maintain a problem size which can fit into the U280's 8GB of HBM. The number of CUs was limited to 4 in the case of PW advection because of the maximum number of 32 AXI4 ports supported by the the Alveo U280 shell. For this benchmark each compute unit requires 7 ports, corresponding to one per field and an additional port for the small data. The exception to this mapping is with DaCe, where there is no option to replicate the number of CUs and the results that are presented are for 1 CU. The tracer advection kernel was implemented using a single CU, since each of its 17 input arguments was mapped to a separate memory port. Whilst some ports could have been bundled together for the tracer advection benchmark to reduce the number of ports of each CU, for example reducing to 12 ports for the input and output fields plus one bundled port for the rest of the arguments would allow for 2 CUs. However this bundling would affect performance and heuristics would likely be required by our transformation to identify when to combine bundles.

Our stencil flow leverages the original PSyclone based Fortran code. Furthermore, the codes were ported to C/C++ to be compatible with SODA-opt and Vitis HLS and Python to be compatible with DaCe. All approaches used the AMD Xilinx Vitis HLS as the backend. Since Stencil-HMLS and SODA-opt generate LLVM-IR that is provided to the AMD Xilinx backend, they had to be compiled with -O0 in Vitis, as otherwise important optimisations are removed such as the copying of small data into local memory in our approach, or the II becomes large. DaCe, by comparison generates HLS C/C++ code that is used as input to the AMD Xilinx Vitis HLS frontend.

The connectivity to HBM was done manually for our approach, SODA-opt and AMD Xilinx Vitis. DaCe generates the connectivity file automatically along side the host code. However, it does not support automatic multi-bank assignment and thus the largest problem size that we measure across the technologies for the PW advection kernel can not be handled using DaCE. Supporting this would require manual construction of the SDFG, which is beyond the automatic optimisations that are part of the comparison in this section. StencilFlow presents these same limitations, as it is built atop of DaCe.

The runtime numbers for StencilFlow could not be obtained and will not be discussed in the following. Whilst PW advection could be compiled for sizes 8M and 32M they did not complete their execution under 10 minutes, a likely indicator of deadlock, and tracer advection could not be expressed in StencilFlow due to the lack of support for subselections which are a core part of this benchmark. Although it needs further development to overcome the technical challenges presented by these benchmarks, the tool is promising, reaching an II of 1 thanks to the optimisations undertaken.

\begin{figure}
	\begin{subfigure}{0.49\columnwidth}
		\includegraphics[width=\columnwidth]{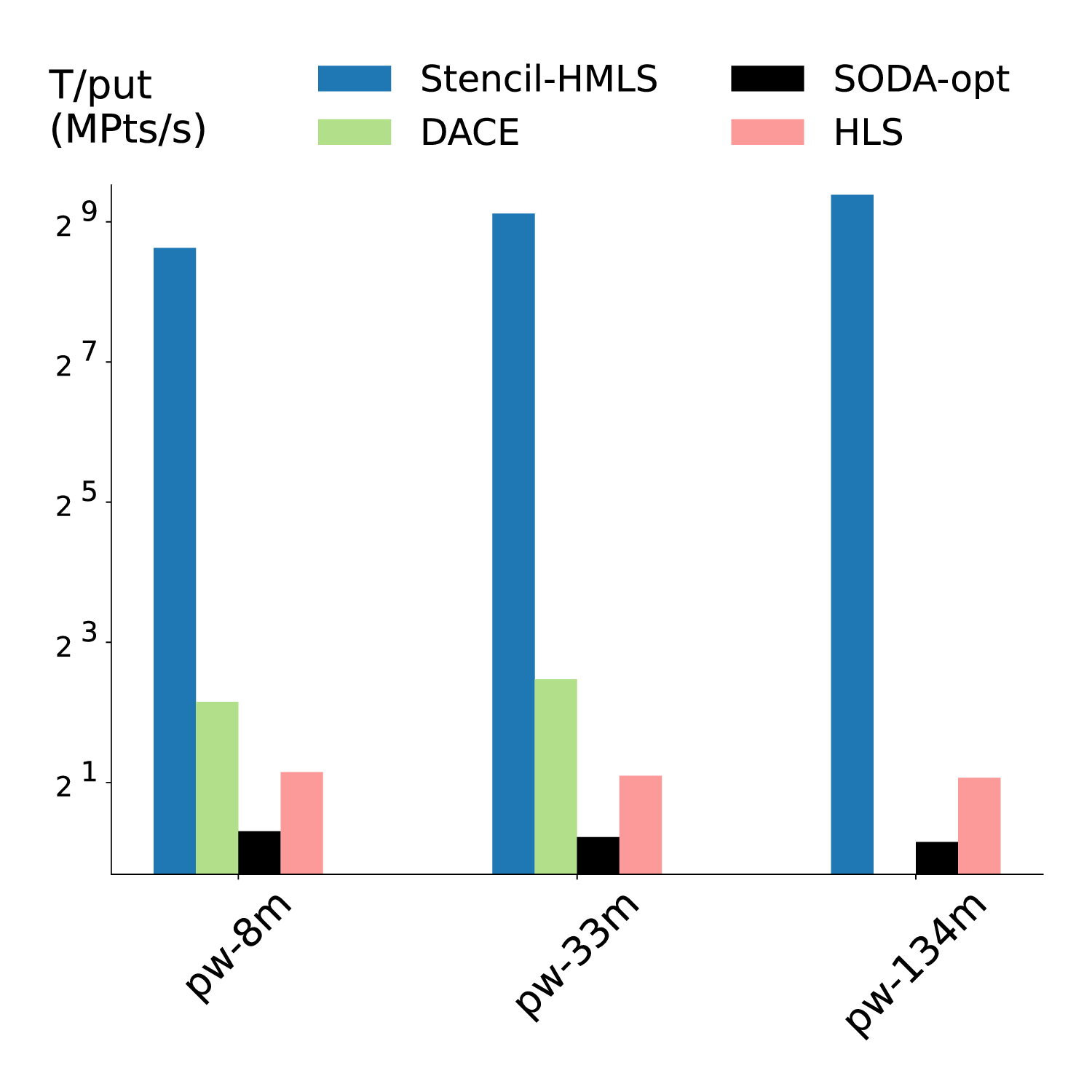}
	\end{subfigure}
	\begin{subfigure}{0.49\columnwidth}
		\includegraphics[width=\columnwidth]{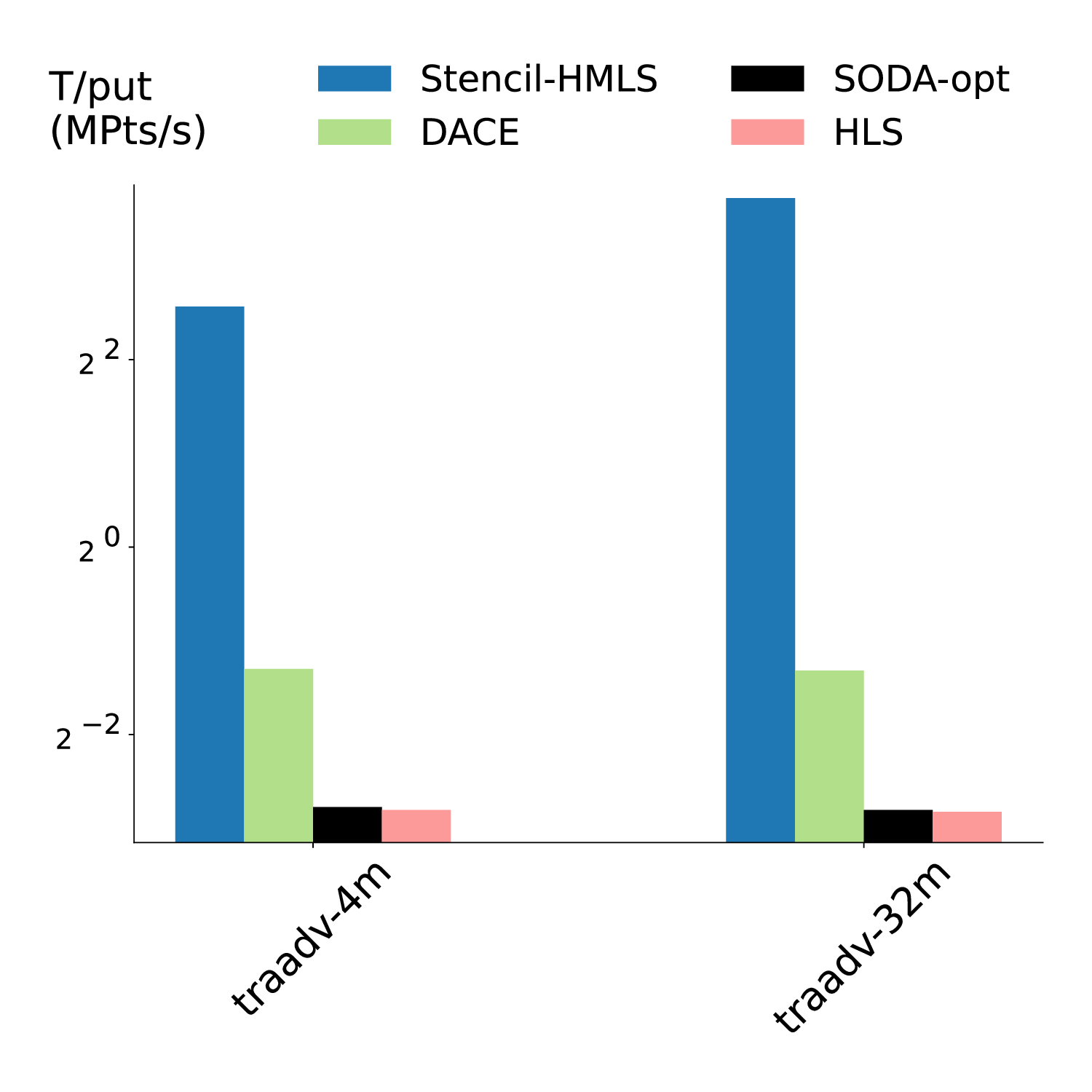}
	\end{subfigure}
	\caption{Performance comparison of the PW advection and tracer advection kernels run across the different frameworks in MPt/s (higher is better).}
	\label{fig:performance}
\end{figure}

Figure \ref{fig:performance} reports the performance of the other four FPGA programming frameworks using our two benchmark kernels on the U280 FPGA. Note that the numbers for the largest size in PW advection are missing for DaCe since it fails to compile with this framework. Stencil-HMLS delivers much higher performance than the others due the stencil specific optimisation that were discussed in Section \ref{sec:stencil-to-hls}, where the II obtained is 1. Our approach is 90 and 100 times faster than next highest performing framework, DaCe, for the PW advection kernel when run with domain sizes of 8M and 32M respectively. Our approach is also between 14 and 21 times faster than DaCe for the tracer advection benchmark kernel for sizes 8M and 33M. 

This performance difference between our approach and DaCe is partially explained by the DaCe generated code having an II of 9. Furthermore, our transformation undertakes an optimisation which splits the computation per field into separate dataflow stages which improve the overall concurrency. Thus when considering the performance difference we can express this as our approach improving by $4 (CUs) \times 9 (1/9 \; of \; DACE's \; II) \times 3 (split) = 108$, which roughly approximates the advantage seen in Figure \ref{fig:performance}. The dependencies between the stencil computations in tracer advection do not allow for a clean split across components as it is applied for PW advection, which explain the reduced advantage of Stencil-HMLS over the other frameworks.

SODA-opt delivers the lowest overall performance with the PW advection benchmark, and this is because loop unrolling had to be disabled as otherwise the pipeline that was generated is too large to fit within the U280 FPGA's resources, even when the number of full unrolls is set to 1. Moreover, the memory buffers generated by SODA-opt were disabled, as they were translated into \texttt{malloc} calls in the IR, which is incompatible with the AMD Xilinx HLS backend. Performance of SODA-opt and unoptimised Vitis HLS code for the tracer advection benchmark is comparable, where SODA-opt achieves an II of 164 and Vitis HLS of 163 on their critical path.

\begin{figure}
	\centering
	\begin{subfigure}{0.49\columnwidth}
		\includegraphics[width=\columnwidth]{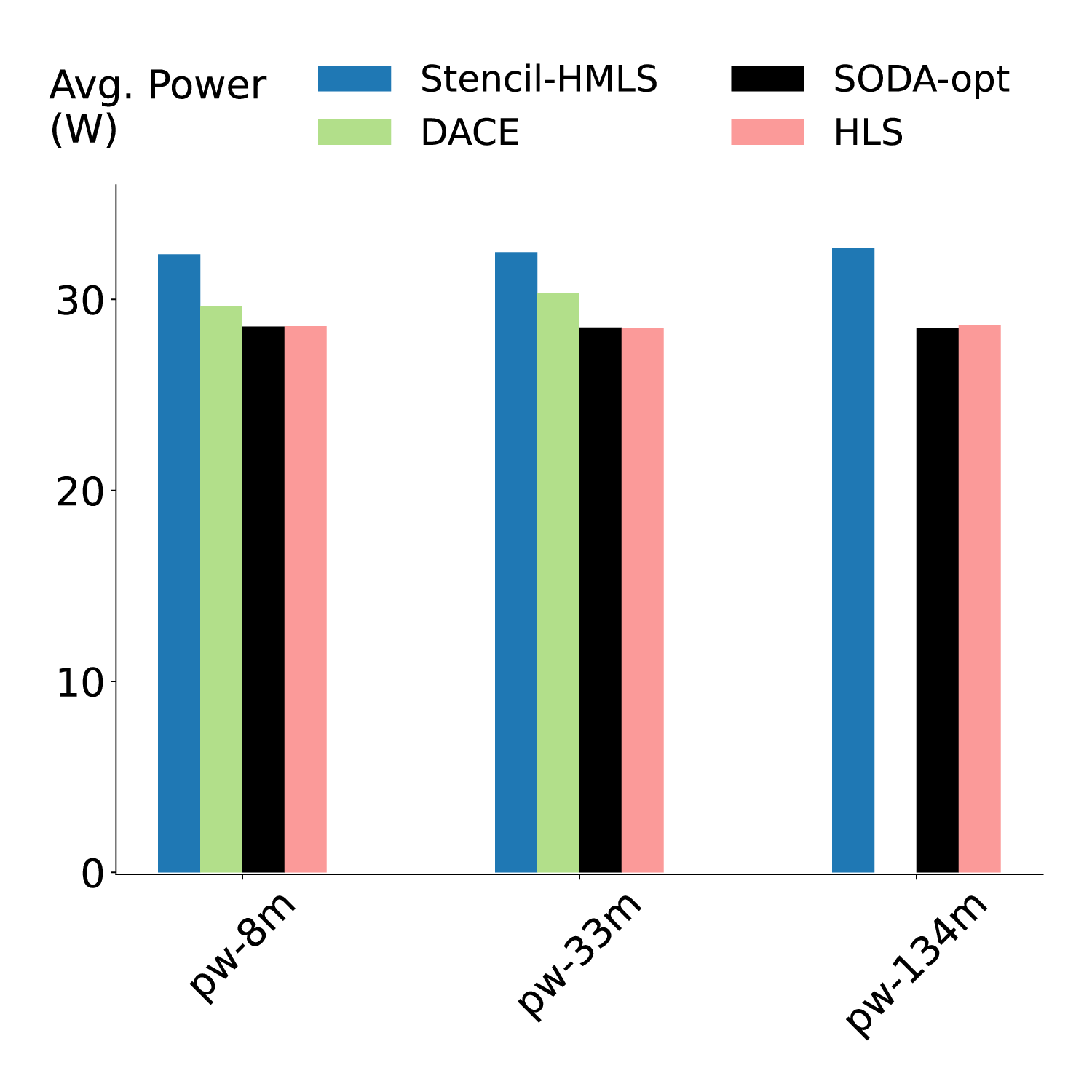}
	\end{subfigure}
	\begin{subfigure}{0.49\columnwidth}
		\includegraphics[width=\columnwidth]{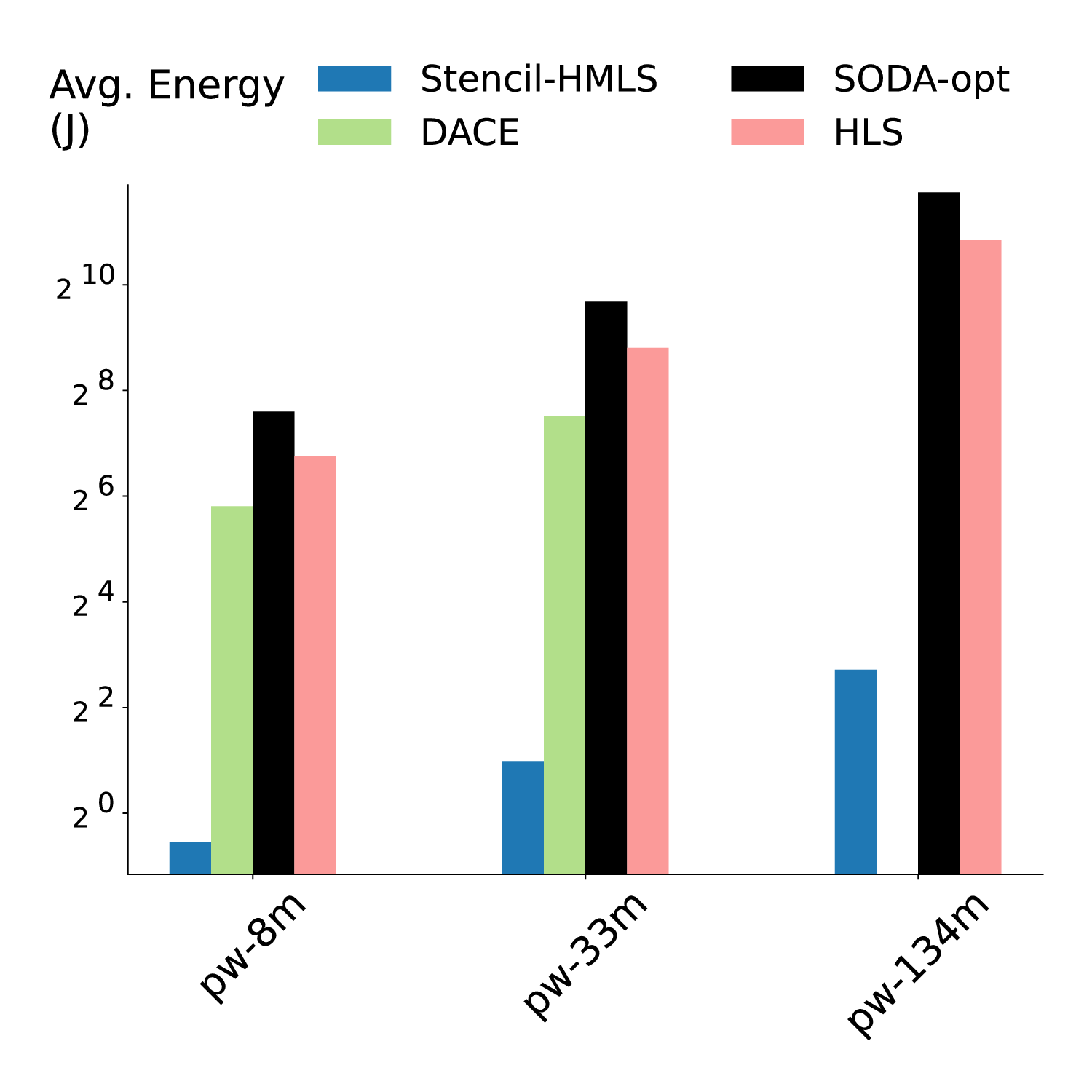}
	\end{subfigure}
	\caption{Average power draw and average energy consumption of PW advection across the different frameworks (lower is better).}
	\label{fig:power-pw_advection}
\end{figure}

\begin{figure}
	\centering
	\begin{subfigure}{0.49\columnwidth}
		\includegraphics[width=\columnwidth]{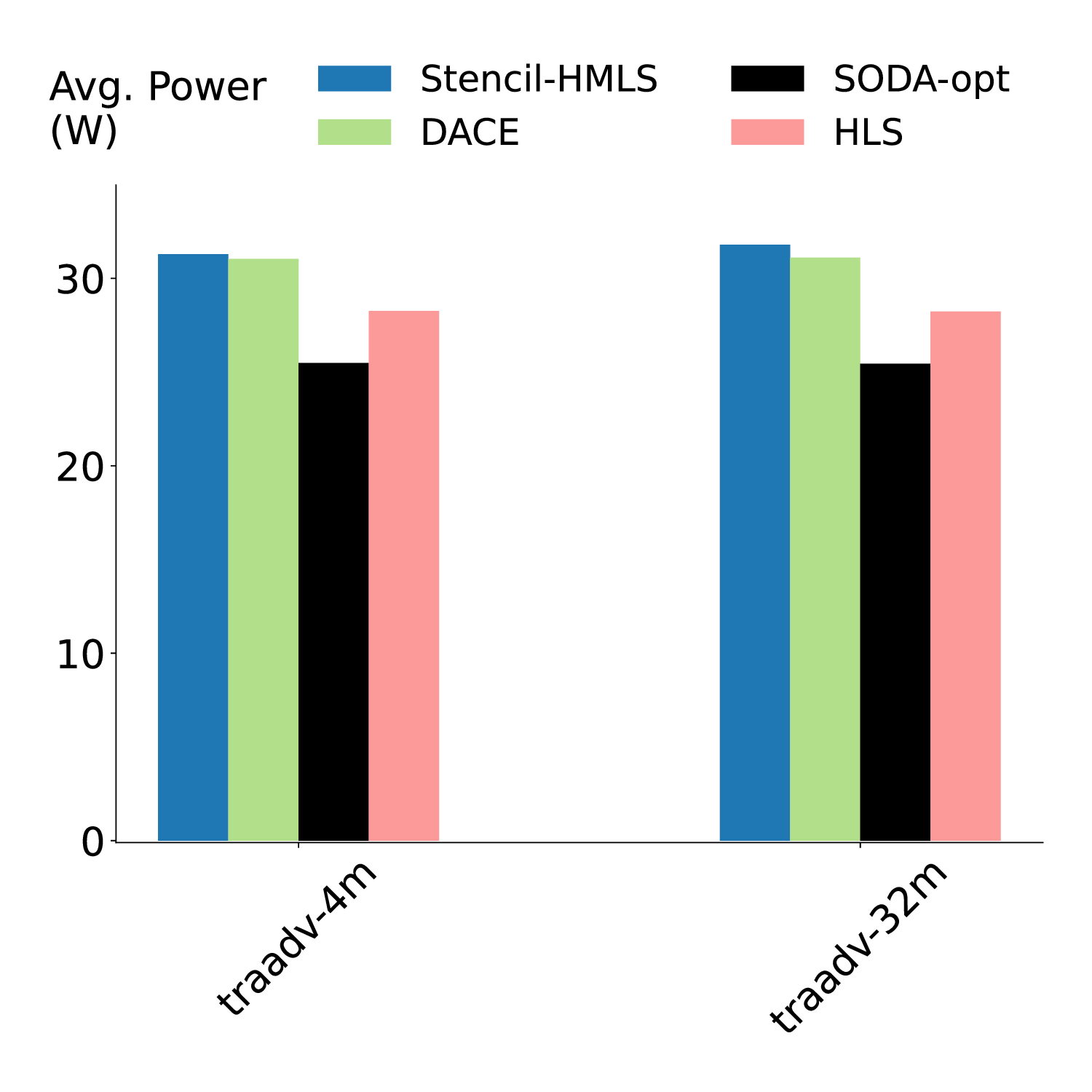}
	\end{subfigure}
	\begin{subfigure}{0.49\columnwidth}
		\includegraphics[width=\columnwidth]{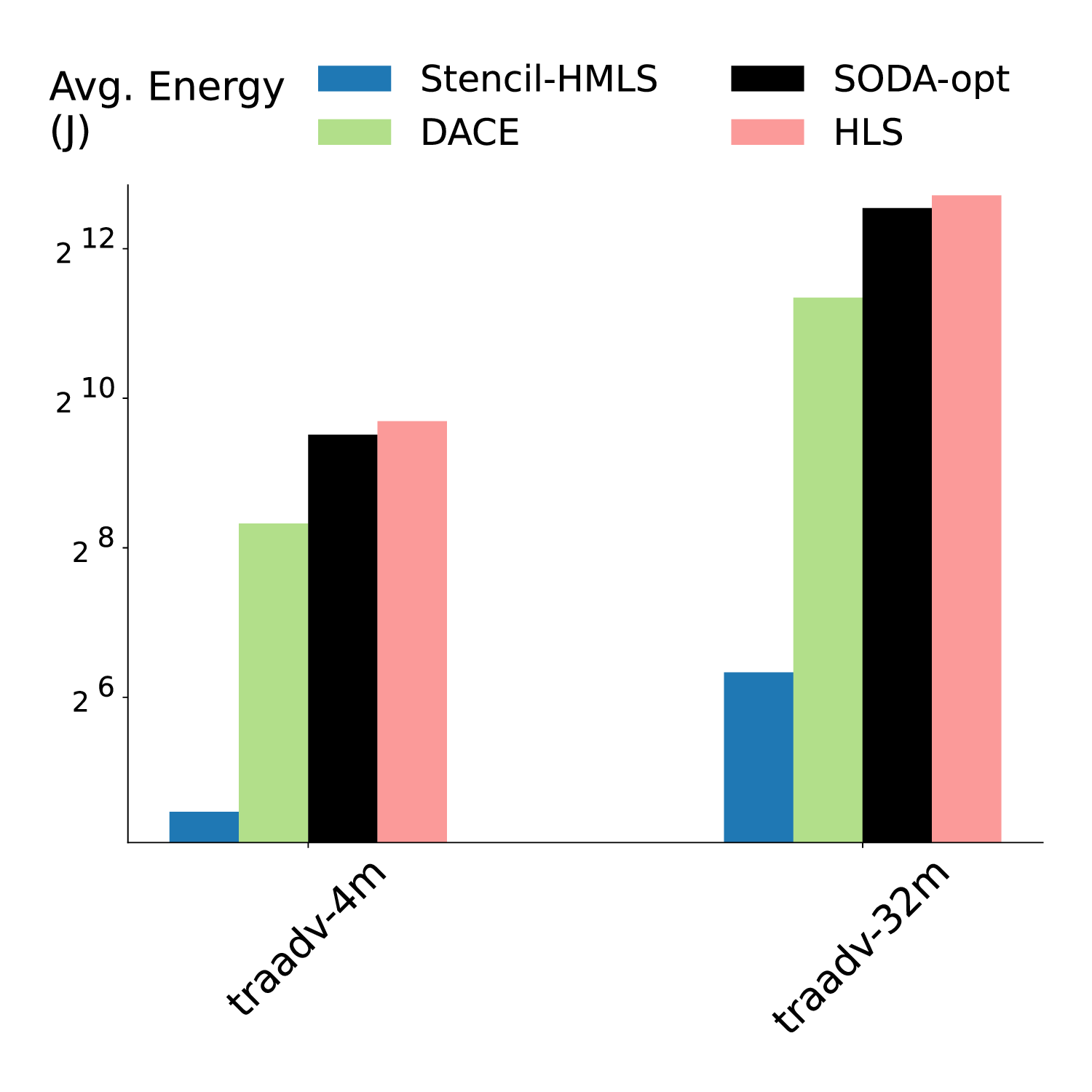}
	\end{subfigure}
	\caption{Average power draw and average energy consumption of tracer advection across the different frameworks (lower is better).}
	\label{fig:power-tra_adv}
\end{figure}

Figure \ref{fig:power-pw_advection} and \ref{fig:power-tra_adv} present the power draw and energy efficiency for the PW advection and tracer advection benchmarks respectively. For both benchmarks it can be seen that our approach is the most energy efficient. Whilst the power draw of our approach is marginally greater than the other frameworks, the fact that FPGA kernels built using our approach run for much less time means that in total we consume 85 and 92 times less energy for sizes 8M and 32M for PW advection, and 14 and 22 times less energy for sizes 8M and 33M of tracer advection than DACE, which is the next most energy efficient. Similarly, the longer run times of both SODA-opt and Vitis HLS lead to significantly higher energy consumption. However, their power draw is, for both kernels, the lowest, SODA-opt drawing the least power for the tracer advection kernel.

Tables \ref{tab:res_usage-pw_advection} and \ref{tab:res_usage-tra_adv} respectively present the resource utilisation for the PW advection and tracer advection benchmarks. For both kernels, there is roughly no variation in resource utilisation for Vitis HLS, since there are no local arrays of size dependent of the problem size. However, we see variations in all other frameworks. In the case of our approach, this is due to the copies of the small data areas into local memory, whose size effectively grow with the problem size. We expected the other frameworks to perform similar optimisations and surprisingly, we see no variation in SODA-opt for PW advection.

Finally, it should be noted that we could expect the results for SODA-opt to be different than reflected in this experimentation if the Bambu HLS backend had been used instead of Xilinx AMD Vitis HLS. Even though the experimentation in \cite{agostini2022mlir} shows that the performance with of AMD Xilinx's backend and Bambu HLS is similar, SODA-opt has progressed in the direction of a better integration with Bambu HLS ever since. We noticed that with Bambu HLS it was not necessary to remove their internal buffers, as their backend could manage \texttt{malloc}. Unfortunately, the shell in the U280 used for our experimentation was not supported by Bambu HLS.

\newcommand{\e}[1]{\times 10^{#1}}


\begin{table}[]
    \centering
    \caption{Resource usage for the PW advection kernel}
    \label{tab:res_usage-pw_advection}
    \begin{tabular}{c|c|S[table-format = 2.2]|S[table-format = 2.2]|S[table-format = 2.2]|S[table-format = 2.2]}
    \toprule
    \textbf{FRAMEWORK} & \textbf{SIZE} & \textbf{\%LUTs} & \textbf{\%FFs} & \textbf{\%BRAM} & \textbf{\%DSPs} \\
	\midrule
	              & \textbf{8M} & 4.301515709376534 & 3.0229043937162494 & 14.285714285714285 & 1.3076241134751774 \\
	\textbf{Stencil-HMLS} & \textbf{32M} & 4.312561364752086 & 3.0280436917034854 & 14.484126984126986 & 1.3076241134751774 \\
				  & \textbf{134M} & 4.332965144820815 & 3.0310735763377514 & 14.087301587301587 & 1.3076241134751774 \\
	\midrule
    \textbf{DACE} & \textbf{8M} & 8.352509818360334 & 1.995543384879725 & 5.5059523809523805 & 0.4875886524822695 \\
				  & \textbf{32M} & 8.35726558664703 & 1.997384327442317 & 5.5059523809523805 & 0.4875886524822695 \\
	\midrule
				  & \textbf{8M} & 0.8192194403534611 & 0.5132010922925871 & 0.0992063492063492 & 0.15514184397163122 \\
	\textbf{SODA-opt} & \textbf{32M} & 0.8213672066764851 & 0.5139297987236131 & 0.0992063492063492 & 0.15514184397163122 \\
				  & \textbf{134M} & 0.8242053264604812 & 0.5148119170348552 & 0.0992063492063492 & 0.15514184397163122 \\
	\midrule
				  & \textbf{8M} & 1.0995029455081002 & 0.5161926239567992 & 0.0992063492063492 & 0.12189716312056738 \\
	\textbf{HLS} & \textbf{32M} & 1.1023410652920962 & 0.5169596833578792 & 0.0992063492063492 & 0.12189716312056738 \\
              & \textbf{134M} & 1.1066365979381443 & 0.5178418016691212 & 0.0992063492063492 & 0.12189716312056738 \\
	\midrule
    \textbf{StencilFlow} & \textbf{8M} & 4.799413966617575 & 3.056616654393716 & 16.865079365079367 & 3.667996453900709 \\
              & \textbf{32M} & 4.8107664457535595 & 3.0705387825233186 & 16.865079365079367 & 3.667996453900709 \\
    \bottomrule
    \end{tabular}
\end{table}

\begin{table}[]
    \centering
    \caption{Resource usage for the tracer advection kernel}
    \label{tab:res_usage-tra_adv}
    \begin{tabular}{c|c|S[table-format = 2.2]|S[table-format = 2.2]|S[table-format = 2.2]|S[table-format = 2.2]}
    \toprule
      \textbf{FRAMEWORK}   & \textbf{SIZE} & \textbf{\%LUTs} & \textbf{\%FFs} & \textbf{\%BRAM} & \textbf{\%DSPs} \\
     \midrule
	\textbf{Stencil-HMLS} & \textbf{8M} & 27.052344133529697 & 18.868702442317133 & 62.74801587301587 & 4.122340425531915 \\
				  & \textbf{33M} & 27.137641138929798 & 18.902184585174275 & 62.74801587301587 & 4.122340425531915 \\
	\midrule
	\textbf{DACE} & \textbf{8M} & 11.470836401570937 & 3.6513561610211096 & 10.069444444444445 & 0.6759751773049646 \\
				  & \textbf{33M} & 11.517473613156602 & 3.670647704958272 & 10.069444444444445 & 0.7092198581560284 \\
	\midrule
	\textbf{SODA-opt} & \textbf{8M} & 14.814294918998527 & 2.790600454099166 & 0.744047619047619 & 0.24379432624113476 \\
				  & \textbf{33M} & 14.772566887579774 & 2.7983861070201277 & 0.744047619047619 & 0.24379432624113476 \\
	\midrule
	\textbf{HLS} & \textbf{8M} & 13.998067010309278 & 2.49942470544919 & 0.744047619047619 & 0.24379432624113476 \\
				  & \textbf{33M} & 14.020004909180168 & 2.500421882670594 & 0.744047619047619 & 0.24379432624113476 \\
    \bottomrule
    \end{tabular}
\end{table}

\section{Conclusions and further work} \label{sect:conclusions}
This paper presented a multi-layered approach to automatic stencil optimisation based on MLIR. It extends the xDSL framework with the new HLS dialect that acts as a high-level abstractions for powerful concepts for FPGA performance such as task-level parallelism and pipelining. Here we present the direct lowering to LLVM IR used to synthesise the FPGA bitstream through AMD Xilinx Vitis HLS, but it opens up alternative paths for further optimisation such as lowering to CIRCT. We describe the lowering of the cornerstone of stencil optimisation in xDSL, the \textit{stencil} dialect, into the \textit{HLS} dialect and how this is further optimised through the FPGA-specific transformation of shift buffers. We demonstrate through our experimentation with two real-life case studies that our approach is both performant and energy efficient compared to the state-of-the-art.

Future work includes but is not limited to:
\begin{itemize}
	\item Explore opportunities for increased performance and/or energy efficiency gains through the lowering of the \textit{HLS} dialect to CIRCT.
	\item Switch to a dynamic shape implementation for the \textit{stencil} dialect for FPGA. The current implementation with static shape needs of the generation of a new bitstream per problem size. Whilst this is not a problem with other architectures, it is time consuming with FPGAs.
	\item Study the performance of the optimisation when the number of memory ports is not a limiting factor. This could be tested on the AMD Xilinx VCK5000, for example, where this limitation does not exist. This would enable further replication of the kernel to better utilise the area of the device.
\end{itemize}

\section*{Acknowledgment}
The authors acknowledge EPCC at the University of Edinburgh and EPSRC who have funded this work (EP/T517884/1), the ExCALIBUR xDSL project (EP/W007940/1), and the ExCALIBUR H\&ES FPGA testbed (ST/W000334/1) and AMD Xilinx HACC program for access to compute resource used in this work. On a similar note, the authors would like to thank Tiziano De Matteis and Alexandros Nikolaos Ziogas for their help setting up DACE for the evaluation of this work, Nicolas Bohm Agostini for his help setting up SODA-opt, Johannes de Fine Licht and Andreas Kuster for their support in the evaluation of StencilFlow, and Mark Klaisoongnoen for his assistance collecting power and energy numbers from the Alveo U280.

\newpage
\appendix

\section{Artifact Description Appendix}
\label{sec:ad}

\subsection{Description}

\subsubsection{Check-list (artifact meta information)}

{\small
\begin{itemize}  
  \item {\bf Program: } Python 3
  \item {\bf Data set: }Runs were based on the tiny, small, medium, large, and huge problem sizes which are described in this paper with a standard execution of the benchmark provided by the official STAC-A2 reference implementation. 
  \item {\bf Data set: }Runs were based on the 8M, 32M and 134M points for PW advection and 8M and 33M for tracer advection.
  \item {\bf Run-time environment: } A machines running Linux was used, equipped with an Alveo U280 the environment at the time of writing was used (XRT 202110.2.11.634, xdma and xdma-dev 201920.3 deployment and development target platforms).
  \item {\bf Hardware: } We used a Xilinx Alveo U280, hosted in a systems with a 24-core Intel Xeon Platinum 8260M with 512GB of RAM. 
  \item {\bf Binary: } AMD Xilinx Vitis is required to synthesise the kernel and generate the bitstream. We built our bitstreams with Vitis 2021.2.
  \item {\bf Execution: } We built and executed all executables on Linux.
  \item {\bf Output: } We measured all performance using OpenCL's performance mechanism and also undertook secondary timing using the \emph{omp\_getwtime} call with microsecond resolution to ensure consistency. 
  \item {\bf Publicly available?: }Yes, except for f++, which is still undergoing a licensing process and is expected to be released open source in the near future in \url{https://fpga.epcc.ed.ac.uk/community/fortran.html}. To mitigate this, our Makefiles have a target to generate LLVM IR from xDSL and another to pass output of f++ that we effectively provide to Vitis.
\end{itemize}
}

\subsubsection{Hardware dependencies}
Any machine running Linux with appropriate Alveo U280 card installed.
\subsubsection{Software dependencies}
Python 3.10 or higher, GCC 10.2.0 (other versions will work as well).
\subsubsection{Datasets}
Runs were based on the PW advection and tracer advection benchmarks using the problem sizes described in this paper.
\subsection{Installation}
Figure \ref{fig:artifact_tree} shows the structure of the artifact relevant for its evaluation. The four frameworks evaluated in this work, xDSL, DACE and SODA-opt are provided in their respective directories, cloned as \emph{git submodules}. To install xDSL or DACE:
\begin{enumerate}
	\item{Access \path{xdsl} or \path{dace}}
	\item{Run \emph{pip install .}}
\end{enumerate}

The instructions to install SODA-opt are more complicated and are provided by the original authors in a README under \path{soda-opt}.
All the bitstreams are ultimately synthesised using AMD Xilinx Vitis HLS after the corresponding lowerings to LLVM IR or HLS, depending on the tool. The host codes are written in OpenCL. To launch the bitstreams on the FPGA it is sufficient to compile and synthesise the host codes. Makefile are provided for all the source codes.

\subsection{Experiment workflow}
\begin{enumerate}
\item Port the chosen kernels to the input languages of the tools: Python for DACE, C for HLS and SODA-opt and Fortran for Stencil-HMLS if we are using the PSyclone front-end.
\item For Stencil-HMLS feed the code to xDSL and then the resulting LLVM IR to f++ to adapt it to the requirements of the AMD Xilinx Vitis HLS backend. For DACE, qualify the kernel function with the \emph{@dace.program} decorator and apply the FPGATransformSDFG to transform the SDFG into a version amenable to FPGA. For SODA-opt run the source code through \emph{cgeist}, from \path{soda-opt/polygeist} and then call SODA-opt through a script like the one provided in \path{H2RC/benchmarks/soda-opt/pw_advection/256x256x32/run-outline-affine_for-opt_full-vitishls.sh}. For HLS, simply compile and link with \emph{v++}.
\item Compile the host OpenCL code using GCC.
\item Execute the host code, which will launch the bitstream.
\end{enumerate}


\section{Artifact Evaluation}

\begin{figure}
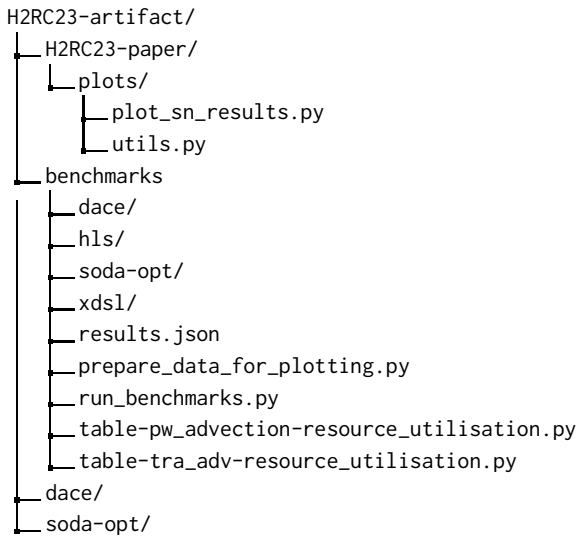

\centering
\begin{subfigure}{\columnwidth}
\dirtree{%
.1 H2RC23-artifact/.
.2 H2RC23-paper/.
.3 plots/.
.4 plot\_sn\_results.py.
.4 utils.py.
.2 benchmarks.
.3 dace/.
.3 hls/.
.3 soda-opt/.
.3 xdsl/.
.3 results.json.
.3 prepare\_data\_for\_plotting.py.
.3 run\_benchmarks.py.
.3 table-pw\_advection-resource\_utilisation.py.
.3 table-tra\_adv-resource\_utilisation.py.
.2 dace/.
.2 soda-opt/.
}
\end{subfigure}
\caption{Relevant directories of the artifact for evaluation.}
\label{fig:artifact_tree}
\end{figure}

The artifact of this paper can be found in \url{https://zenodo.org/record/8272497}. The structure of the directory tree of the artifact relevant for its evaluation is shown in Figure \ref{fig:artifact_tree}. The four frameworks evaluated in Section \ref{sect:evaluation} are cloned in the top level as submodules in the directories \path{dace}, \path{xdsl} (Stencil-HMLS builds atop of xDSL) and \path{soda-opt}. The bitstreams used to generate the results presented in the paper are provided with the artifact, although Makefiles are also provided for their generation. They are found under the corresponding directory for each framework under \path{benchmarks/<framework>/<size>}, except for DACE, for which the bitstreams are found under \path{benchmarks/dace/<size>/.dacecache/<kernel>/build}. Similarly, for all the frameworks except DACE the host codes are found under \path{benchmarks/<framework>} along side a Makefile to build them. They are run with \emph{host <size>}, where size is 0, 1 or 2 (only for PW advection), for each of the sizes presented in the paper for each problem, 0 being the smallest. For DACE the host codes are found under \path{benchmarks/dace/<size>/.dacecache/<kernel>/} and is simply run with \emph{host}.  Alternatively, the results can be generated from scratch with the script \path{benchmarks/run_benchmarks.py}.

The plots in Section \ref{sect:evaluation} can be generated from the data in \path{benchmarks/results.json}. To do so:
\begin{enumerate}
\item Access \path{H2RC23-paper/plots}.
\item Run \path{plot\_sn\_results.py}.
\end{enumerate}

The bitstreams are generated differently for each framework. For Stencil-HMLS, for PW advection the \emph{all-xdsl} target in the Makefile for each size lowers the code to LLVM IR through xDSL. The \emph{vitis} target synthesises the LLVM IR pre-adapted by f++ and provided in the artifact. There are similar targets for tracer advection. DACE provides its own Makefiles with the target \emph{hw} under \path{benchmarks/dace/<size>/.dacecache/<kernel>/build}. In the case of SODA-opt, we provide Makefiles with a single target that take care of the whole process. HLS also has its own Makefile.

\bibliography{references}

\begin{thebibliography}{10}

\bibitem{agostini2022mlir}
Nicolas~Bohm Agostini, Serena Curzel, Vinay Amatya, Cheng Tan, Marco Minutoli, Vito~Giovanni Castellana, Joseph Manzano, David Kaeli, and Antonino Tumeo.
\newblock An mlir-based compiler flow for system-level design and hardware acceleration.
\newblock In {\em Proceedings of the 41st IEEE/ACM International Conference on Computer-Aided Design}, pages 1--9, 2022.

\bibitem{agostini2022soda}
Nicolas~Bohm Agostini, Serena Curzel, David Kaeli, and Antonino Tumeo.
\newblock Soda-opt an mlir based flow for co-design and high-level synthesis.
\newblock In {\em Proceedings of the 19th ACM International Conference on Computing Frontiers}, pages 201--202, 2022.

\bibitem{ben2019stateful}
Tal Ben-Nun, Johannes de~Fine~Licht, Alexandros~N Ziogas, Timo Schneider, and Torsten Hoefler.
\newblock Stateful dataflow multigraphs: A data-centric model for performance portability on heterogeneous architectures.
\newblock In {\em Proceedings of the International Conference for High Performance Computing, Networking, Storage and Analysis}, pages 1--14, 2019.

\bibitem{brown2021accelerating}
Nick Brown.
\newblock Accelerating advection for atmospheric modelling on xilinx and intel fpgas.
\newblock In {\em 2021 IEEE International Conference on Cluster Computing (CLUSTER)}, pages 767--774. IEEE, 2021.

\bibitem{brown2021porting}
Nick Brown.
\newblock Porting incompressible flow matrix assembly to fpgas for accelerating hpc engineering simulations.
\newblock In {\em 2021 IEEE/ACM International Workshop on Heterogeneous High-performance Reconfigurable Computing (H2RC)}, pages 9--20. IEEE, 2021.

\bibitem{brown2019s}
Nick Brown and David Dolman.
\newblock It's all about data movement: Optimising fpga data access to boost performance.
\newblock In {\em 2019 IEEE/ACM International Workshop on Heterogeneous High-performance Reconfigurable Computing (H2RC)}, pages 1--10. IEEE, 2019.

\bibitem{de2018designing}
Johannes de~Fine~Licht, Michaela Blott, and Torsten Hoefler.
\newblock Designing scalable fpga architectures using high-level synthesis.
\newblock In {\em Proceedings of the 23rd ACM SIGPLAN Symposium on Principles and Practice of Parallel Programming}, pages 403--404, 2018.

\bibitem{de2021stencilflow}
Johannes de~Fine~Licht, Andreas Kuster, Tiziano De~Matteis, Tal Ben-Nun, Dominic Hofer, and Torsten Hoefler.
\newblock Stencilflow: Mapping large stencil programs to distributed spatial computing systems.
\newblock In {\em 2021 IEEE/ACM International Symposium on Code Generation and Optimization (CGO)}, pages 315--326. IEEE, 2021.

\bibitem{eldridge2021mlir}
Schuyler Eldridge, Prithayan Barua, Aliaksei Chapyzhenka, Adam Izraelevitz, Jack Koenig, Chris Lattner, Andrew Lenharth, George Leontiev, Fabian Schuiki, Ram Sunder, et~al.
\newblock Mlir as hardware compiler infrastructure.
\newblock In {\em Workshop on Open-Source EDA Technology (WOSET)}, 2021.

\bibitem{ferrandi2021bambu}
Fabrizio Ferrandi, Vito~Giovanni Castellana, Serena Curzel, Pietro Fezzardi, Michele Fiorito, Marco Lattuada, Marco Minutoli, Christian Pilato, and Antonino Tumeo.
\newblock Bambu: an open-source research framework for the high-level synthesis of complex applications.
\newblock In {\em 2021 58th ACM/IEEE Design Automation Conference (DAC)}, pages 1327--1330. IEEE, 2021.

\bibitem{karp2021high}
Martin Karp, Artur Podobas, Niclas Jansson, Tobias Kenter, Christian Plessl, Philipp Schlatter, and Stefano Markidis.
\newblock High-performance spectral element methods on field-programmable gate arrays: implementation, evaluation, and future projection.
\newblock In {\em 2021 IEEE International Parallel and Distributed Processing Symposium (IPDPS)}, pages 1077--1086. IEEE, 2021.

\bibitem{klaisoongnoen2022fast}
Mark Klaisoongnoen, Nick Brown, and Oliver~Thomson Brown.
\newblock Fast and energy-efficient derivatives risk analysis: Streaming option greeks on xilinx and intel fpgas.
\newblock In {\em 2022 IEEE/ACM International Workshop on Heterogeneous High-performance Reconfigurable Computing (H2RC)}, pages 18--27. IEEE, 2022.

\bibitem{klaisoongnoen2022low}
Mark Klaisoongnoen, Nick Brown, and Oliver~Thomson Brown.
\newblock Low-power option greeks: Efficiency-driven market risk analysis using fpgas.
\newblock In {\em International Symposium on Highly-Efficient Accelerators and Reconfigurable Technologies}, pages 95--101, 2022.

\bibitem{piacsek1970conservation}
Steve~A Piacsek and Gareth~P Williams.
\newblock Conservation properties of convection difference schemes.
\newblock {\em Journal of Computational Physics}, 6(3):392--405, 1970.

\bibitem{fortran-fpga}
Gabriel Rodriguez-Canal, Nick Brown, Tim Dykes, Jess Jones, and Utz-Uwe Haus.
\newblock Fortran high-level synthesis: Reducing the barriers to accelerating hpc codes on fpgas.
\newblock In {\em 33rd International Conference on Field-Programmable Logic and Applications}, 2023.

\bibitem{psyclone_bench}
Science and Technology Facilities~Council (STFC).
\newblock Psyclonebench: Small benchmarks used to inform the development of the psyclone domain-specific compiler, 2021.

\bibitem{wang2015hardware}
Chao Wang, Junneng Zhang, Xi~Li, Aili Wang, and Xuehai Zhou.
\newblock Hardware implementation on fpga for task-level parallel dataflow execution engine.
\newblock {\em IEEE Transactions on Parallel and Distributed Systems}, 27(8):2303--2315, 2015.

\bibitem{ye2021scalehls}
Hanchen Ye, Cong Hao, Hyunmin Jeong, Jack Huang, and Deming Chen.
\newblock Scalehls: Achieving scalable high-level synthesis through mlir.
\newblock In {\em Proceedings of the Workshop on Languages, Tools, and Techniques for Accelerator Design (LATTE’21)}, 2021.

\bibitem{zhang2007multiwindow}
Hui Zhang, Mingxin Xia, and Guangshu Hu.
\newblock A multiwindow partial buffering scheme for fpga-based 2-d convolvers.
\newblock {\em IEEE Transactions on Circuits and Systems II: Express Briefs}, 54(2):200--204, 2007.

\end{thebibliography}

\end{document}